\documentclass{article}
\usepackage{ijcai25}

\usepackage{times}
\usepackage{soul}
\usepackage{url}
\usepackage[hidelinks]{hyperref}
\usepackage[utf8]{inputenc}
\usepackage[small]{caption}
\usepackage{graphicx}
\usepackage{amsmath}
\usepackage{amsthm}
\usepackage{amsfonts}
\usepackage{booktabs}
\usepackage{algorithm}
\usepackage{algorithmic}
\usepackage[switch]{lineno}
\usepackage{subfig}
\usepackage{xcolor} 
\usepackage{array} 
\usepackage{multicol}
\usepackage{tabularx}
\usepackage{algorithm}
\usepackage{algorithmic}
\usepackage{multirow}
\usepackage{adjustbox}
\usepackage{flushend}
\usepackage{subfig}

\title{Phenotypic Profile-Informed Generation of Drug-Like Molecules via Dual-Channel Variational Autoencoders}

\author{
Hui Liu
\and
Shiye Tian\and
Xuejun Liu$^{*}$\\
\affiliations
College of Computer and Information Engineering, Nanjing Tech University, Nanjing, 211800, China.\\
\emails
\{hliu, xjliu\}@njtech.edu.cn,
}

\begin{document}

\maketitle

\begin{abstract}
The de novo generation of drug-like molecules capable of inducing desirable phenotypic changes is receiving increasing attention. However, previous methods predominantly rely on expression profiles to guide molecule generation, but overlook the perturbative effect of the molecules on cellular contexts. To overcome this limitation, we propose SmilesGEN, a novel generative model based on variational autoencoder (VAE) architecture to generate molecules with potential therapeutic effects. SmilesGEN integrates a pre-trained drug VAE (SmilesNet) with an expression profile VAE (ProfileNet), jointly modeling the interplay between drug perturbations and transcriptional responses in a common latent space. Specifically, ProfileNet is imposed to reconstruct pre-treatment expression profiles when eliminating drug-induced perturbations in the latent space, while SmilesNet is informed by desired expression profiles to generate drug-like molecules. Our empirical experiments  demonstrate that SmilesGEN outperforms current state-of-the-art models in generating molecules with higher degree of validity, uniqueness, novelty, as well as higher Tanimoto similarity to known ligands targeting the relevant proteins. Moreover, we evaluate SmilesGEN for scaffold-based molecule optimization and generation of therapeutic agents, and confirmed its superior performance in generating molecules with higher similarity to approved drugs. SmilesGEN establishes a robust framework that leverages gene signatures to generate drug-like molecules that hold promising potential to induce desirable cellular phenotypic changes. The source code and datasets are available at: \url{https://github.com/hliulab/SmilesGEN}.
\end{abstract}

\section{Introduction}
In the realm of drug discovery, target-based drug design (TDD) has emerged as the dominant strategy over the past few decades, driven by advancements in molecular biology and proteomics~\cite{swinney2011were,attwood2021trends}. TDD is centered on the identification and validation of specific molecular targets-typically proteins or RNAs-that play a critical role in disease processes~\cite{munson2024novo}. Once a target is confirmed, compounds are designed or screened based on their ability to interact with the target in a manner that modulates its activity, thereby leading to therapeutic benefits~\cite{meissner2022emerging}. However, the process of molecular target validation is inherently intricate and resource-intensive, which result in a narrow focus on well-characterized targets~\cite{vincent2022phenotypic}. Additionally, TDD is vulnerable to off-target effects, where drugs interact with unintended targets, potentially causing unpredictable and sometimes severe side effects~\cite{sadri2023target}. 

In contrast, phenotypic drug discovery (PDD) adopts a different approach by concentrating on the observable effects of compounds on disease-relevant biological systems~\cite{berg2021future,vincent2022phenotypic,moffat2017opportunities}. In fact, PDD has a long and established history, with many new drugs identified based on the observed therapeutic effects on disease phenotypes~\cite{vincent2022phenotypic}. A major advantage of PDD is that it does not require prior knowledge of specific drug targets, making it particularly advantageous in cases where the underlying molecular mechanism of a disease is not well understood~\cite{moffat2017opportunities,meissner2022emerging}. However, the absence of a well-defined molecular target in phenotypic drug discovery poses significant challenges on refining the chemical structure of lead compounds to optimize their efficacy, safety, and drug-like properties. A promising strategy to overcome these challenges involves leveraging transcriptional response as indicators of phenotypic changes to guide the drug design and optimization process~\cite{malandraki2021use}. Expression profiles provide a detailed, high-resolution snapshot of how a compound influences cellular pathways and processes, offering a molecule-level understanding of the observed phenotypic outcomes. By comparing the expression profiles of treated and untreated cells, researchers can identify potential drug candidates that elicit desired phenotypic shifts. Therefore, expression profiles can serve as phenotypic signatures in the discovery process. Thereafter, the terms ``expression profile" and ``phenotypic profile" are used interchangeably in this study.

Several large-scale assays have been conducted to measure a wealth of drug-induced transcriptional responses \textit{in vitro}. For example, the L1000 platform~\cite{SUBRAMANIAN20171437} offers a cost-effective, high-throughput method for generating expression profiles subjected to drug perturbations. Based on such data resource, deep generative models, such as generative adversarial networks (GANs)~\cite{Goodfellow2014GenerativeAN} and variational autoencoders (VAEs)~\cite{2013arXiv1312.6114K}, have been proposed to design molecules with potential to elicit intended bioactivity~\cite{malandraki2021use}. However, existing approaches generate molecules without considering the cellular context relevant to specific disease. To guide drug discovery using phenotypic profile, several methods have been developed to utilize drug-induced expression profiles as conditional variables for molecular generation. These methods include ExpressionGAN~\cite{mendez2020novo}, TRIOMPHE~\cite{kaitoh2021triomphe}, GxVAEs~\cite{li2024gxvaes}, and Gex2SGen~\cite{das2023gex2sgen}. Despite their advancement, these phenotype-based methods have two primary limitations. First, they rely exclusively on post-treatment expression profiles to guide drug generation, neglecting the change of expression profiles induced by drug perturbations. Second, they do not model the interplay between molecules and cellular environment, which is crucial for generating molecules with potential to induce desired phenotypic changes.

To address these issues, we propose SmilesGEN, a novel deep generative framework designed for \textit{de novo} generation of drug-like molecules. SmilesGEN uniquely integrates drug-induced gene expression changes with phenotype-informed molecule generation within a unified framework. Unlike previous approaches, SmilesGEN uses dual-channel variational autoencoders to explicitly model the interplay between molecules and cellular environment. By mapping both molecules and expression profiles to a common latent space, the drug-induced phenotypic changes can be straightforwardly formulated. SmilesGEN is designed to satisfy two critical constraints. First, by mitigating the disturbative effects of molecules on post-treatment phenotypic profiles, SmilesGEN aims to restore the pre-treatment cellular state. Second, when conditioned on desired phenotypic profiles, SmilesGEN is engineered to generate molecules with potential to induce such phenotypic outcomes. Extensive experiments demonstrate that SmilesGEN outperforms current state-of-the-art methods, generating molecules with potential bioactivity and desirable drug-like properties. By establishing the link between molecule generation and cellular context, SmilesGEN effectively increases the likelihood of generating molecules that interact productively with disease-associated cellular machinery.

The main contributions of this study are summarized as below:
\begin{itemize}
    \item \textbf{Phenotype-informed \textit{de novo} molecule generation}: Unlike previous methods designed to generate molecules with proper physicochemical property, this study aims to generate molecules that potentially elicit desirable phenotypic outcomes.
    \item \textbf{Explicit formalization of drug perturbative effect on phenotypic change}: this study explicitly formalizes the phenotypic change to capture the net effect of the drug perturbation, enabling to extract more informative features for molecular generation.
    \item \textbf{Superior performance over state-of-the-art (SOTA) methods}: Extensive experiments demonstrate that our method outperforms two baselines and five previous methods, and generates therapeutic molecules with chemical structures highly similar to those of known ligands.
\end{itemize}

 \section{Related Work}
\subsection{De Novo Drug Generation}
Some generative models have been employed to design novel molecular structures, aiming to explore chemical space and generate molecules with both novelty and chemical validity. These approaches often utilize Simplified Molecular Input Line Entry System (SMILES) strings as input, which are well-suited for generative models like recurrent neural networks~\cite{segler2018generating,merk2018novo}, VAEs~\cite{gomez2018automatic,blaschke2018application,zhang2022novo}, and Transformers~\cite{wang2022transformer,dollar2021attention}. Attention-based models leverage attention mechanisms to learn the ``molecular grammar" from  SMILES strings, enabling the generation of coherent molecular sequences. Despite the advantages of compact and standardized representation provided by SMILES, these methods often produce invalid molecular structures, as the stringed representation of SMILES does not always guarantee chemical validity. So, some methods employ Graph Neural Networks (GNNs) to represent and generate molecular structures~\cite{de2018molgan,li2018learning,jin2020hierarchical}, in order to capture the chemical structure of molecules. In addition, reinforcement learning techniques have been incorporated into some methods to optimize the structures of generated molecules~\cite{zhavoronkov2019deep,popova2018deep}.

\subsection{Drug Generation from Expression Profile}
 With the advancement of RNA sequencing technology, the generation of large-scale multi-omics data has enabled researchers to integrate omics data into generative models for the design of de novo molecules. For instance, TRIOMPHE~\cite{kaitoh2021triomphe} employed a variational autoencoder framework conditioned on e n profiles to generate molecules. Gex2SGen~\cite{das2023gex2sgen} and GxVAEs~\cite{li2024gxvaes} followed similar ideas that used the desired expression profiles as input to design drug-like molecules capable of eliciting phenotypic changes. Mendez-Lucio et al.~\cite{mendez2020novo} further employed a generative adversarial network (GAN) in conjunction with expression profiles to automatically design molecules with a high probability to induce therapeutic effects. Although the integration of transcriptomic profiles into generative models marks a significant advancement, these methods overlook the interplay between drug molecules and cellular context. 

\begin{figure*}[!htbp]
    \centering
    \includegraphics[width=\textwidth]{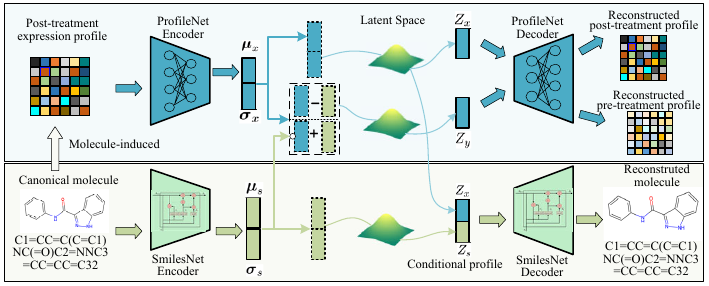}
    \caption{SmilesGEN framework comprises two interacting variational autoencoders, SmilesNet and ProfileNet, which work jointly to extract features from molecules and phenotypic profiles and reconstruct molecules and post-treatment signatures, as well as pre-treatment signatures. 
    }  \label{fig:framework}
\end{figure*} 

\section{Methods}
\subsection{SimlesGEN Framework}
The proposed SmilesGEN framework comprises dual-channel VAEs: a molecule VAE (SmilesNet) and an expression profile VAE (ProfileNet). As shown in Figure~
\ref{fig:framework}, SmilesNet is a VAE ~\cite{Chung2014EmpiricalEO} that utilizes gated recurrent unit (GRU) backbone for the encoding and generation of molecules. SmilesNet is pretrained on a large-scale SMILES descriptors, enabling it to explore the extensive chemical space and generate meaningful molecular structures. ProfileNet is another VAE composed of multiple feed-forward layers to encode and decode expression profiles. A key feature of ProfileNet is to explicitly formalize the net effect of drug perturbations on cellular state in the latent space, by enforcing the constraint that the encoded post-treatment expression profile, when adjusted by the molecular representation, should restore the corresponding pre-treatment expression profile. Conceptually, this formalization simulates the therapeutic effects in a reverse direction that removal of drug treatment allows the cellular environment to revert to its unperturbed state. By modeling the dynamic interplay between molecules and cellular environments, our model enhances its capacity to learn informative and expressive features related to both molecule structures and their induced phenotypic changes. Once trained, the encoded expression profiles are fed into the SmilesNet decoder to generate molecules with a high likelihood of eliciting the desired phenotypic outcomes. 

\subsection{Drug Variational Autoencoder}
The molecules are represented using SMILES strings, a compact and standardized notation for describing molecular structures. Let a  molecule $S$ be denoted by a string $S=[s_1,s_2,\cdots,s_n]$, in which each character corresponds to an atom or a chemical bond. Suppose the posterior distribution learned by SmilesNet encoder is denoted as $q_{\phi}(Z_s|S)$ that is approximate to a normal distribution $\mathcal{N}(\boldsymbol{\mu}_s, \boldsymbol{\sigma}_s)$, where $\boldsymbol{\mu}_s$ and $\boldsymbol{\sigma}_s$ denote the mean and standard deviation. A latent variable $Z_s$ is sampled from $q_{\phi}(Z_s|S)$ and then used to reconstruct the molecule
$S$ following a stochastic process $S\sim p_{\varphi}(S|Z)$ parameterized by $\varphi$:
\begin{equation}
    p_{\varphi}(S|Z_s,Z_x)=\prod_{i=1}^n p_{\varphi}(s_i|s_{1:i-1},Z_s,Z_x).
\end{equation}
in which $Z_x$ represents the conditional variable drawn from the probability distribution learned from drug-induced phenotypic profiles (see below). During the pre-training phase, $Z_x$ in is substituted with $Z_s$. SmilesNet tries to maximize the lower bound of the true log-marginal likelihood as follows:
\begin{equation}
\begin{aligned} 
\mathcal{L}_{S}(\phi,\varphi)=&\mathbb{E}_{Z_s\sim q_{\phi}(Z_s|S)}(\log p_{\varphi}(S|Z_s,Z_x))\\ & -D_{KL}(q_{\phi}(Z_s|S)||p_{\varphi}(Z_s)) 
\end{aligned}
\end{equation}
where $\mathbb{E}(\cdot)$ is an expectation operation, $\phi$ and $\varphi$ denote the parameters of the SmilesNet encoder and decoder, respectively.

\subsection{Profile Variational Autoencoder}
Denote the post-treatment expression profile by $X=(x_1,x_2,\cdots,x_m)$, where each element represents the expression levels of a gene when subjected to a molecule. Similarly, denote by $Y=(y_1,y_2,\cdots,y_m)$ the pre-treatment expression profile of corresponding genes. Suppose the posterior distribution learned by ProfileNet encoder is denoted as $q_{\theta}(Z_x|X)$ that is approximate to a normal distribution $\mathcal{N}(\boldsymbol{\mu}_x, \boldsymbol{\sigma}_x)$, where $\boldsymbol{\mu}_x$ and $\boldsymbol{\sigma}_x$ denote the mean and standard deviation. From the posterior distribution, a latent variable $Z_x$ is sampled and taken as the conditional variable by ProfileNet decoder to reconstruct $X$ following a generative process $X \sim p_{\psi}(X|Z_x)$. ProfileNet aims to maximize the marginal likelihood to generate the expression profile from the generative process:
\begin{equation}
\begin{split}  
\mathcal{L}_X(\theta,\psi)=\mathbb{E}_{Z_x\sim q_{\theta}(Z_x|X)}(\log p_{\psi}(X|Z_x)) 
\\- D_{\text{KL}}(q_{\theta}(Z_x|X)||p_{\psi}(Z_x))
\end{split}
\end{equation}
where $\theta$ and $\psi$ are actually the parameters of the ProfileNet encoder and decoder, respectively. 

\subsection{Modeling Drug-Induced Phenotypic Change}
We aim to explicitly formalize the impact of drugs on cellular states within the latent space. Following the assumption of linear additivity that has been widely used in modeling cellular response to drug perturbations~\cite{hetzel2022predicting,huang2024predicting}, we assume that the linear subtraction of two random variables capture the data distribution of the pre-treatment expression profiles (control state), when mitigate the effect of drug perturbation from post-treatment expression profiles. As a result, we can draw samples from the resulting random variable to reconstruct the pre-treatment expression profile using the ProfileNet decoder. Denote the variable $Z_y$ is approximated by the reparameterization trick:
\begin{equation}
Z_y= (\boldsymbol{\mu}_x-\boldsymbol{\mu}_s)+(\boldsymbol{\sigma}_x+\boldsymbol{\sigma}_s)\boldsymbol{\epsilon},
\end{equation}
in which $\boldsymbol{\epsilon}$ is sampled from a standard normal distribution $\mathcal{N}(\mathbf{0},\boldsymbol{I})$. Following the same generative process $p_{\psi}(\cdot)$, we expect to reconstruct the pre-treatment expression profile $Y$ when conditioned on $Z_y$. So, we define another loss function as below:
\begin{equation}
\begin{split}  
\mathcal{L}_Y(\theta,\psi)=\mathbb{E}_{Z_y \sim \mathcal{N}(\boldsymbol{\mu}_x-\boldsymbol{\mu}_s,\boldsymbol{\sigma}_x+\boldsymbol{\sigma}_s)}(\log p_{\psi}(Y|Z_y)) 
\\- D_{\text{KL}}(q_{\theta}(Z_y|X)||p_{\psi}(Z_y))
\end{split}
\end{equation}

\subsection{Full Objective}
Taken together, we define the total loss function as below:
\begin{equation}
\mathcal{L}(\phi,\varphi,\theta,\psi)=\mathcal{L}_{S}(\phi,\varphi)+\mathcal{L}_{X}(\theta,\psi)+\mathcal{L}_{Y}(\theta,\psi)
\end{equation}
The parameter are optimized by minimizing the total loss function. The training process of SmileGEN consists of two distinct stages. As outlined in Algorithm 1,  the first stage involves pretraining SmilesNet on large-scale dataset of molecules. Once pretraining is completed, the encoder of SmilesNet is frozen. In subsequent training stage, the model takes as input the drug $S$, its perturbed expression profile $X$, and unperturbed expression profile $Y$ to reconstruct them. The parameters are optimized by minimizing the objective loss function. After training, we retain the ProfileNet encoder and SmilesNet decoder to generate drug-like molecules from perturbed expression profiles.

\begin{algorithm}[!htpb]
	\renewcommand{\algorithmicrequire}{\textbf{Data:}}
	\caption{Training Procedure of SmilesGEN.}
	\label{algorithm: masp}
	\begin{algorithmic}[1]
		\REQUIRE
			SMILES string $\mathbf{S}$,
            drug-perturbed profiles $\mathbf{X}$,
			unperturbed profiles $\mathbf{Y}$
        \STATE /*Pre-training of SmilesNet*/
        \FOR {$i \leftarrow$  $1$  $to$  $\mathbf{SEpoch}$}
            \STATE Train SmilesNet to minimize Eq.(2). 
            \STATE Update parameters $\varphi$ and $\phi$.  
        \ENDFOR
        \STATE Freeze SmilesNet encoder parameter $\phi$.
        \STATE /*Joint Training of SmilesNet and ProfileNet*/
        \FOR {$j \leftarrow$  $1$  $to$  $\mathbf{GEpochs}$}
            \STATE Train SmilesNet and ProfileNet to minimize  Eq.(6).  
            \STATE Update parameters $\theta$, $\psi$, and $\varphi$.  
        \ENDFOR
	\end{algorithmic}
\end{algorithm}

In our practice, we employed the bidirectional GRUs as the backbone of SmilesNet encoder and decoder. Both the encoder and decoder were composed of three hidden layers of size 192. The ProfileNet encoder and decoder adopted multiple forward-feed layers with sizes 512, 256, and 192, respectively. During the pretraining stage, SmilesNet was optimized using the Adam optimizer with a learning rate of 5e-4 and a dropout rate of 0.1. The maximum length of the generated SMILES strings was restricted to 100 characters. In the subsequent training stage, both SmilesNet and ProfileNet were optimized using Adam optimizer with a reduced learning rate of 1e-4. The maximum epochs of the pre-training and training stages were set to 100 and 500, respectively. The batch size for both stages was set at 64. SmilesGEN was implemented using PyTorch 2.3.0. The training and all evaluation experiments were conducted on a CentOS Linux 8.2.2004 (Core) system, equipped with a GeForce RTX 4090 GPU and 128GB of memory.

\begin{figure*}[!hptb]
    \centering
    \includegraphics[width=1.0\textwidth]{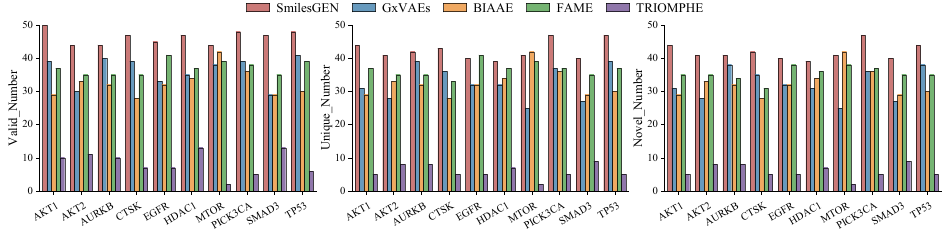}
    \caption{Performance comparison in terms of validity, uniqueness and novelty of generated molecules by SmilesGEN and four competing methods using the expression profiles induced by genetic perturbations.
    }  \label{fig:property}
\end{figure*}

\begin{table*}[!htb]
    \centering
    \newcolumntype{Y}{>{\centeringarraybackslash}X}
    \begin{tabularx}{\textwidth}{c|cccccccc}
        \hline
        Target Protein&Euclidean &Cosine 
& ExpressionGAN & TRIOMPHE & BIAAE & FAME & GxVAEs & SmilesGEN \\
        \hline

AKT1&0.35 &0.18 &0.32 &0.23 &0.43&0.44&0.66 &  $\boldsymbol{0.98}$\\     
AKT2&0.36 &0.33 &0.29 &0.50 &0.42&0.55&0.38 &$\boldsymbol{0.81}$\\  
AURKB&0.29 &0.34 &0.36 &0.25 &0.43&0.55&0.50 &$\boldsymbol{0.63}$\\  
CTSK&0.18 &0.29 &0.31 &0.32 &0.37&0.43&0.41 &$\boldsymbol{0.51}$\\        
EGFR&0.26 &0.26 &0.30 &0.47 &0.47&0.76&0.58 &$\boldsymbol{0.81}$\\
HDAC1&0.39 &0.25 &0.34 &0.24 &0.42&0.46&0.49 &$\boldsymbol{0.56}$\\
MTOR&0.33 &0.19 &0.39 &0.30 &0.49&0.42&0.50 &$\boldsymbol{0.75}$\\
PIK3CA&0.25 &0.24 &0.26 &0.32 &0.35&0.55&0.43 &$\boldsymbol{0.75}$\\
SMAD3&0.44 &0.32 &0.44 &0.38 &0.45&0.52&0.51 &$\boldsymbol{0.76}$\\
TP53&0.40 &0.36 &0.46 &0.59 &0.58&0.73&0.71 &$\boldsymbol{1.00}$\\
        \hline
    \end{tabularx}
\caption{Comparison of highest Tanimoto coefficients achieved the SmilesGEN and seven competing methods} \label{tab:Tanimoto}
\end{table*}

\section{Experiments}
\subsection{Datasets}
For performance evaluation, we used the expression profiles induced by the genetic perturbation of 10 commonly target genes associated with the treatment of prevalent cancers. Among them, eight genes (AKT1, AKT2, AURKB, CTSK, EGFR, HDAC1, MTOR, and PIK3CA) were knockdown, and two genes (SMAD3 and TP53) were over-expressed by genetic perturbations. To evaluate the generated molecules, we obtained inhibitors and agonists that have been confirmed to target these proteins from the Drug Target Commons (DTC) database~\cite{tang2018drug}. We compute the performance metrics between these inhibitors/agonists and generated molecules.

We also utilized the expression profiles of cancer patients from The Cancer Genome Atlas (TCGA) repository~\cite{weinstein2013cancer} to guide the generation of therapeutic molecules. Specifically, we extracted the disease-specific expression profiles of three types of cancers: triple-negative breast cancer (TNBC), colorectal cancer (CRC), and lung adenocarcinoma (LUAD). To evaluate the generated molecules, we retrieved a list of drugs approved for treatment of these cancers from DrugBank database~\cite{wishart2006drugbank}. The generated molecules were then assessed based on their structural similarity to the approved drugs, holding the rationale that structurally similar molecules are likely to exhibit therapeutic effects against the diseases.

\subsection{Evaluation Metrics}
For the generated molecules, we assessed their validity, uniqueness, novelty, and diversity. High validity and uniqueness are indicators of an effective molecule generation process, while high novelty indicates the model is not overfitting to the training data. High diversity reflects the model's ability to generate molecules beyond the molecules in training set. To assess the drug-likeness of the generated molecules, we employed the quantitative estimate of drug-likeness (QED) score. Additionally, we evaluated the similarity between the generated molecules and known ligands of target proteins using Tanimoto similarity based on molecular fingerprints. This multifaceted assessment ensures the effectiveness and applicability of our model in generating molecules with potential therapeutic effect.

\begin{figure*}[!h]
    \centering
        \includegraphics[width=1\textwidth]
        {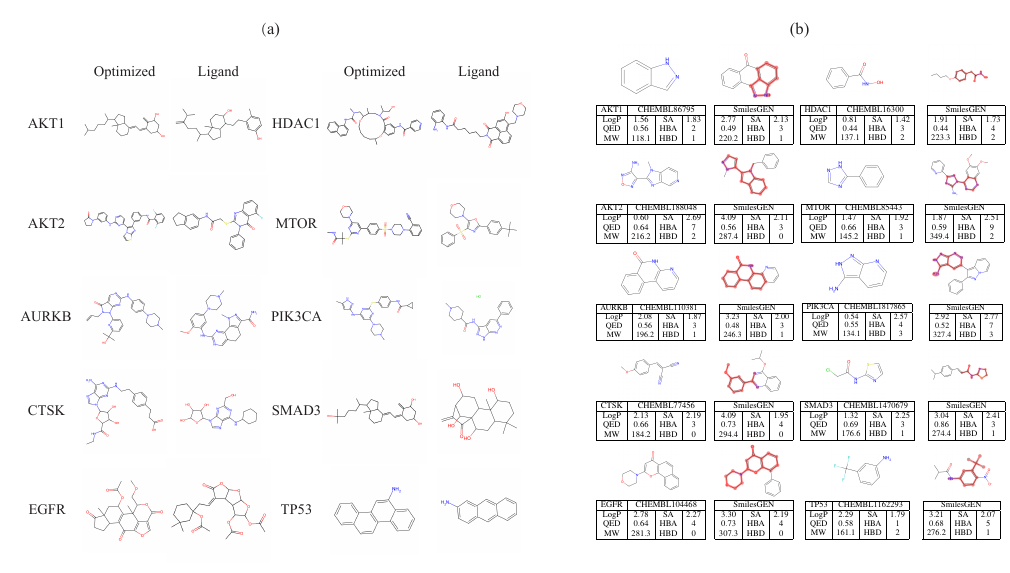}
\caption{Performance evaluation for scaffold-based molecule optimization. (a) Optimized molecules based on benzene ring toward specific gene targets and their nearest known ligands. (b) Comparison between optimized molecules and corresponding short-chain ligands. }  \label{fig:optimize}
\end{figure*}


\subsection{Evaluation on Benchmark dataset}
To validate the model's capability to generate molecules based on desired expression profiles, we conducted experiments using the expression profile induced by genetic perturbations in the MCF7 cell line, a benchmark dataset that has been utilized in previous studies. Specifically, we used the knockdown-induced expression profiles of eight target genes (AKT1, AKT2, AURKB, CTSK, EGFR, HDAC1, MTOR, and PIK3CA) to generate inhibitors, and overexpression-induced profiles of two target genes (SMAD3 and TP53) to generate agonists. We first conducted a comparative analysis of our model against four existing methods in terms of validity, uniqueness, and novelty metrics. The competing methods include GxVAEs~\cite{li2024gxvaes}, BIAAE~\cite{shayakhmetov2020molecular}, FAME~\cite{pham2022fame} and TRIOMPHE~\cite{kaitoh2021triomphe}. As shown in Figure~\ref{fig:property}, the experimental results indicate that the molecules generated by our model achieved higher degree of validity, uniqueness, and novelty than four competing methods.

Next, we evaluated the performance of SmilesGEN against seven existing models by assessing the Tanimoto coefficient between generated molecules and known ligands. To benchmark our method, we included two baseline models based on molecular similarity search (Euclidean and Cosine similarity), alongside five other competing methods: ExpressionGAN~\cite{mendez2020novo}, TRIOMPHE~\cite{kaitoh2021triomphe}, BIAAE~\cite{shayakhmetov2020molecular}, FAME~\cite{pham2022fame}, and GxVAEs~\cite{li2024gxvaes}. For each target gene, 50 molecules were generated using each method, and the Tanimoto coefficients were computed between the generated molecules and the known ligands specific to that target gene. The highest Tanimoto coefficients for each method were reported, as shown in Table~\ref{tab:Tanimoto}. Notably, SmilesGEN consistently achieved significantly higher Tanimoto coefficients compared to all other methods. Although the current state-of-the-art model, GxVAEs, outperformed the other competing methods, its performance remained notably inferior to that of SmilesGEN. Furthermore, Figure~S1 showcases the molecular structures with the highest Tanimoto similarity to known ligands, as generated by SmilesGEN and five competing methods. Such visual comparisons further underscore SmilesGEN’s superior ability to generate molecules with the highly similar structures to known ligands.

Moreover, we further evaluated the drug-likeness and diversity of the molecules generated by SmilesGEN. For each target gene, we generated 1,000 candidate molecules and analyzed their physicochemical properties, including the quantitative estimate of drug-likeness (QED), logarithm of partition coefficient (LogP), molecular weight (MW), hydrogen bond acceptor (HBA), and hydrogen bond donor (HBD). The kernel density distributions of these physicochemical properties for both the generated molecules and the known ligands are presented in Figure~S2. The results revealed that the QED distribution for our generated molecules exhibited a peak around 0.7, while 90\% known ligands displayed peak QED values below 0.55, implying the generated molecules have higher degree of drug-likeness than known ligands. For other physicochemical metrics, the broader distribution range of the generated molecules compared to known ligands further confirmed the better diversity achieved by our method. 

In addition, we applied Lipinski’s Rule of Five (logP$<$10, molecular weight$<$500, hydrogen bond donors$<$5, hydrogen bond acceptors$<$10) to filter the 1,000 generated molecules, and then computed the Tanimoto similarities between the filtered molecules and known ligands. Table~S1 showed the details of generated molecules with Tanimoto similarities exceeding 0.7. It is worth noting that among the refined molecules, more than 50\% of the molecules generated towards desired expression profiles induced by seven gene perturbations (AKT1, AURKB, EGFR, MTOR, PIK3CA, SMAD3, TP53) exhibited high similarity with known ligands (Tanimoto$\ge$0.7). We showcased the molecular structures of the generated molecules with highest Tanimoto similarity to known ligands (Figure~S3). These findings provide strong evidence of our method's capability to generate structurally diverse molecules with high drug-likeness.

\begin{table*}[!htpb]
    \centering
    \caption{Comparison between approved drugs and generated molecules for three types of cancers} \label{tab:patient}
    \setlength{\tabcolsep}{15pt}
    \renewcommand{\arraystretch}{1.1}
    \resizebox{1.0\textwidth}{!}{
    \begin{tabular}{c|cccccc}
        \hline
        Target Protein&
        Approved Drug Name & Approved Molecule & Generated Molecule&Tanimoto \\
        \hline
        TNBC
        &Ribociclib
        & \adjustbox{valign=m}{\includegraphics[width=0.14\linewidth]{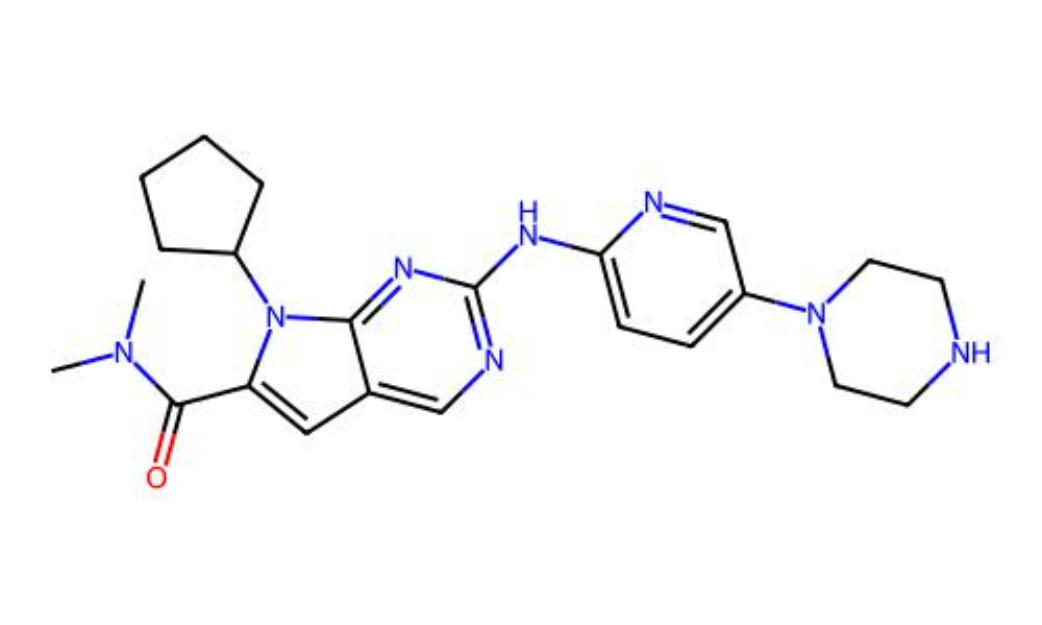}} 
        & \adjustbox{valign=m}{\includegraphics[width=0.14\linewidth]{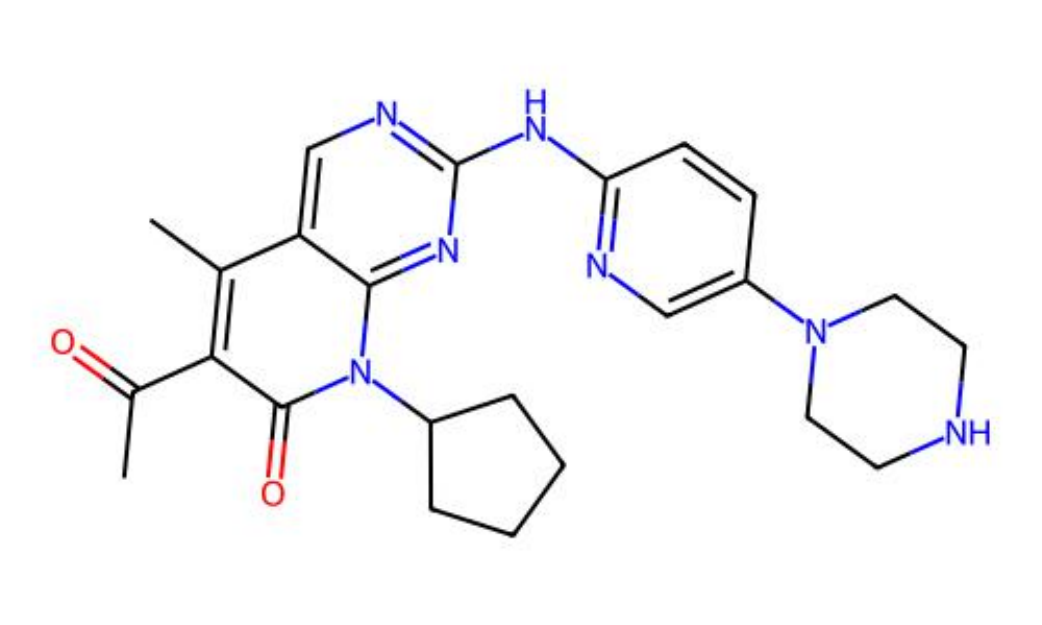}}
        &$\boldsymbol{0.56}$ 
        \\
        \hline
        CRC
        & Cetuximab
        & \adjustbox{valign=m}{\includegraphics[width=0.14\linewidth]{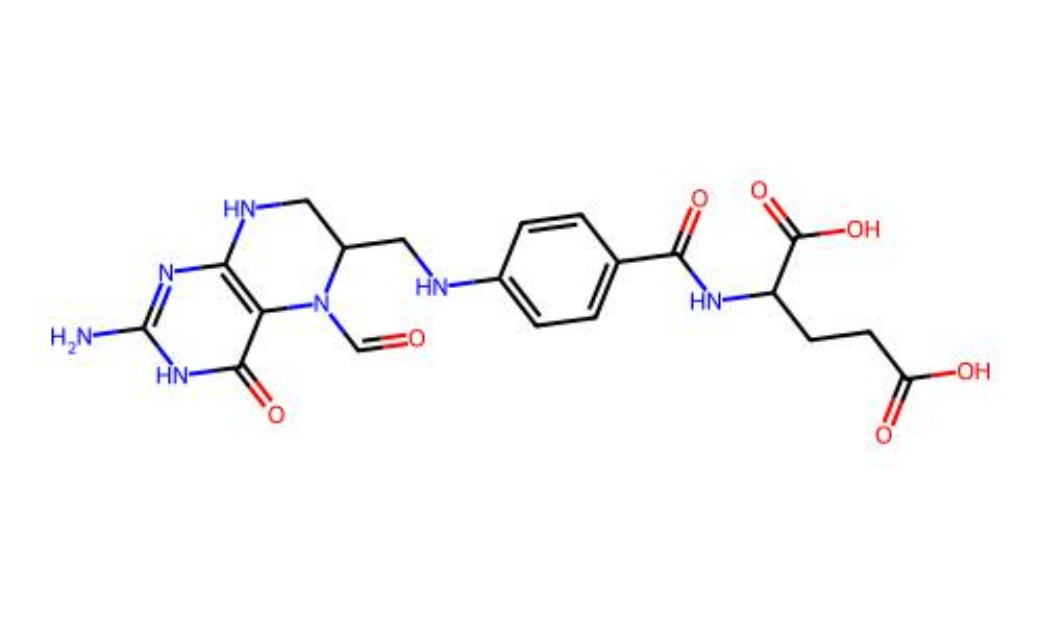}}
        & \adjustbox{valign=m}{\includegraphics[width=0.14\linewidth]{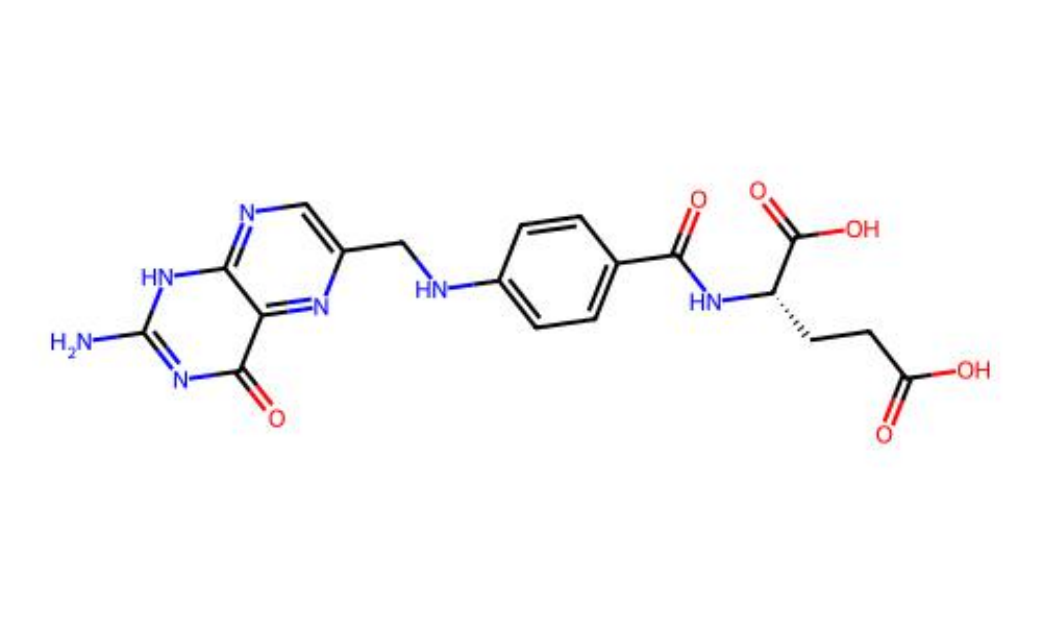}}
        &$\boldsymbol{0.50}$ 
        \\
        \hline
        LUAD
        & Alectinib
        & \adjustbox{valign=m}{\includegraphics[width=0.14\linewidth]{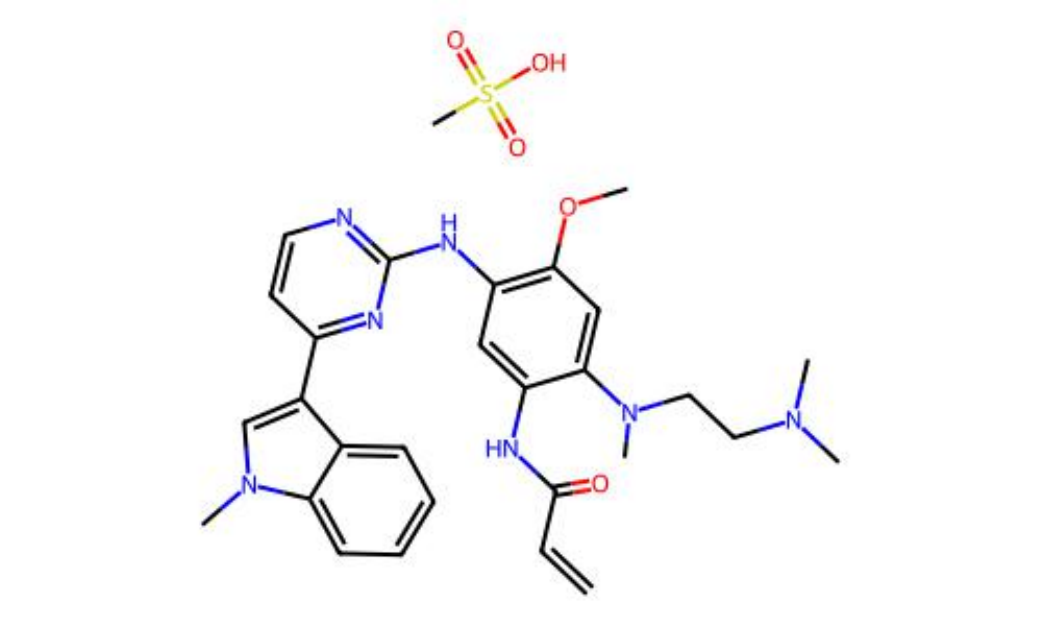}}
        & \adjustbox{valign=m}{\includegraphics[width=0.14\linewidth]{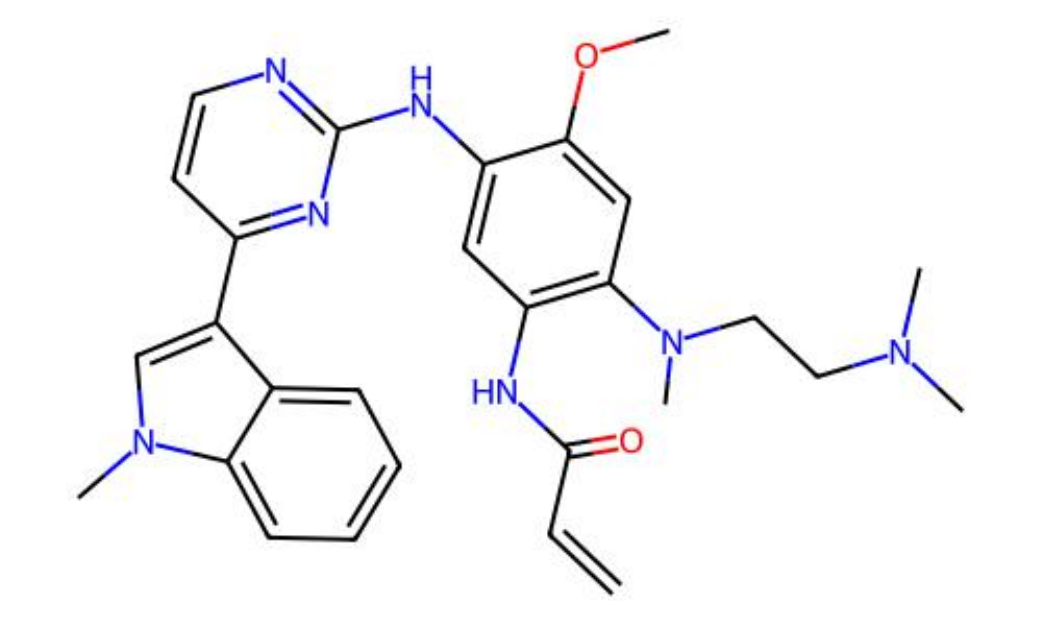}}
        &$\boldsymbol{0.98}$ 
        \\
        \hline

    \end{tabular}
    }
\end{table*}


\subsection{Profile-Based Molecule Optimization}
Molecule optimization is another important task in drug discovery, as it significantly contributes to improving therapeutic efficacy and pharmacokinetic properties. For proof-of-concept purpose, we evaluate our model in optimizing benzene rings—one of the most common scaffolds in drugs—towards active-like molecules, guided by the gene signature induced by genetic perturbations. Specifically, the SMILES of benzene rings and desired expression profiles were taken as input to generate molecular structures. Figure~\ref{fig:optimize}(a) presents the optimized molecules alongside their nearest known ligands. The results demonstrate that the generated molecules not only effectively retained the benzene ring scaffold, but also incorporated functional groups containing atoms such as oxygen (O), nitrogen (N), sulfur (S), and fluorine (F), showcasing the model’s ability to enhance structural diversity and functionality.

Furthermore, we extended the optimization process to short-chain ligands associated with specific target genes. Figure~\ref{fig:optimize}(b) presented the short-chain ligands alongside their optimized counterparts, as well as their physicochemical properties, including LogP, QED, MW, SA, HBA, and HBD. We found that the optimized molecules incorporated additional benzene rings or functional groups, thereby expanding the molecular structures while maintaining or improving their QED values relative to the original short-chain ligands. Meanwhile, other physicochemical properties were enhanced to varying degrees. These findings highlight the strong capability of our model in molecular optimization, offering robust support for drug design and development.



\subsection{Therapeutic Molecule Generation }
We further applied SmilesGEN to design therapeutic molecules for triple-negative breast cancer (TNBC), colorectal cancer (CRC), and lung adenocarcinoma (LUAD). These cancers are all characterized as malignant tumors, and a significant proportion of patients either do not respond to approved drugs or develop resistance over time. We obtained the expression profiles of patients with these cancers from The Cancer Genome Atlas (TCGA). Given the rationale that molecules capable of reversing disease-specific expression profiles might possess therapeutic potential~\cite{tong2023transigen,yamanaka2023novo}. Therefore, we averaged the expression profiles across patients of the same cancer type and inverted the averaged expression profiles by multiplying the values by -1. These disease-specific reversal profiles were subsequently input into SmilesGEN to generate candidate therapeutic molecules.

\begin{figure}[!htbp]
    \centering    
    \includegraphics[width=0.4\textwidth]{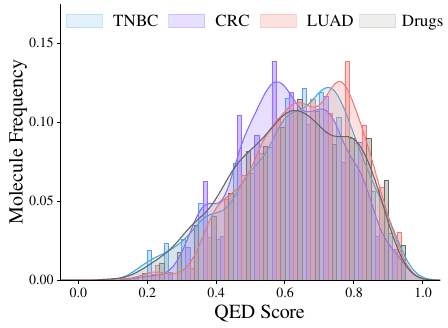}
    \caption{Distribution of QED values of generated molecules using the reversal profiles of three types of cancers. 
    }  \label{fig:QED}
\end{figure}

For each type of cancer, we generated 1,000 candidate molecules. After refining the generated molecules using the Lipinski's Rule of Five, we obtained 783, 859, and 662 molecules for TNBC, CRC, and LUAD, respectively. Figure~\ref{fig:QED} displays the distribution of quantitative estimate of QED values of these molecules compared to approved drugs for treating these diseases. It is noteworthy that the mean QED values of the molecules generated for LUAD and TNBC exceeded 0.75, suggesting that our model is highly effective in generating drug-like molecules with therapeutic potential.

We further calculated the Tanimoto similarity between our generated molecules and known approved drugs. Table~\ref{tab:patient} highlights the molecules generated by our model that exhibit significant similarity to approved drugs for respective cancers. For example, Ribociclib, an approved drug for TNBC treatment~\cite{li2022recent}, is a specific inhibitor of cyclin-dependent kinase 4/6 (CDK4/6), used in the treatment of hormone receptor-positive and human epidermal growth factor receptor 2-negative (HR+/HER2-) advanced or metastatic breast cancer in postmenopausal women. The molecules generated by our model for these three cancers exhibited high Tanimoto similarity to these approved drugs. Notably, when generating therapeutic molecules for lung cancer using the reversal profile, the Tanimoto similarity between the generated molecules and Alectinib Hydrochloride reached 0.98. This high degree of similarity indicates that SmilesGEN effectively captures structural features similar to those of approved drugs, suggesting that our model can generate hit-like molecules with high therapeutic potential. 



\vspace{-0.5cm}

\section{Conclusion}
In this paper, we introduce the SmilesGEN, a novel generative model designed to generate hit-like molecules based on desired phenotypic profiles. Our model explicitly models the interplay between molecules and cellular environments in the latent space, driving the model to learn informative and expressive features of molecules and expression profiles. Experimental results show that SmilesGEN outperforms current state-of-the-art models and can generate hit-like molecules with potential therapeutic effects. We believe that SmilesGEN represents an alternative approach to bridging chemistry and biology in the long and difficult road of drug discovery.

\section*{Acknowledgments}
This work was supported by National Natural Science Foundation of China (No. 62372229), Natural Science Foundation of Jiangsu Province (No. BK20231271).

\bibliographystyle{named}
\bibliography{ijcai25}

\end{document}